\documentstyle[12pt,epsf]{article}
\expandafter\ifx\csname amssym12.def\endcsname\relax \else\endinput\fi
\expandafter\edef\csname amssym12.def\endcsname{%
       \catcode`\noexpand\@=\the\catcode`\@\space}
\catcode`\@=11
\def\undefine#1{\let#1\undefined}
\def\newsymbol#1#2#3#4#5{\let\next@\relax
 \ifnum#2=\@ne\let\next@\msafam@\else
 \ifnum#2=\tw@\let\next@\msbfam@\fi\fi
 \mathchardef#1="#3\next@#4#5}
\def\mathhexbox@#1#2#3{\relax
 \ifmmode\mathpalette{}{\m@th\mathchar"#1#2#3}%
 \else\leavevmode\hbox{$\m@th\mathchar"#1#2#3$}\fi}
\def\hexnumber@#1{\ifcase#1 0\or 1\or 2\or 3\or 4\or 5\or 6\or 7\or 8\or
 9\or A\or B\or C\or D\or E\or F\fi}
\font\tenmsa=msam10 scaled\magstep1
\font\sevenmsa=msam7 scaled\magstep1
\font\fivemsa=msam5 scaled\magstep1
\newfam\msafam
\textfont\msafam=\tenmsa
\scriptfont\msafam=\sevenmsa
\scriptscriptfont\msafam=\fivemsa
\edef\msafam@{\hexnumber@\msafam}
\mathchardef\dabar@"0\msafam@39
\def\dashrightarrow{\mathrel{\dabar@\dabar@\mathchar"0\msafam@4B}}
\def\dashleftarrow{\mathrel{\mathchar"0\msafam@4C\dabar@\dabar@}}

\def\ulcorner{\delimiter"4\msafam@70\msafam@70 }
\def\urcorner{\delimiter"5\msafam@71\msafam@71 }
\def\llcorner{\delimiter"4\msafam@78\msafam@78 }
\def\lrcorner{\delimiter"5\msafam@79\msafam@79 }
\def\yen{{\mathhexbox@\msafam@55 }}
\def\checkmark{{\mathhexbox@\msafam@58 }}
\def\circledR{{\mathhexbox@\msafam@72 }}
\def\maltese{{\mathhexbox@\msafam@7A }}
\font\tenmsb=msbm10 scaled\magstep1
\font\sevenmsb=msbm7 scaled\magstep1
\font\fivemsb=msbm5 scaled\magstep1
\newfam\msbfam
\textfont\msbfam=\tenmsb
\scriptfont\msbfam=\sevenmsb
\scriptscriptfont\msbfam=\fivemsb
\edef\msbfam@{\hexnumber@\msbfam}

\def\widehat#1{\setbox\z@\hbox{$\m@th#1$}%
 \ifdim\wd\z@>\tw@ em\mathaccent"0\msbfam@5B{#1}%
 \else\mathaccent"0362{#1}\fi}
\def\widetilde#1{\setbox\z@\hbox{$\m@th#1$}%
 \ifdim\wd\z@>\tw@ em\mathaccent"0\msbfam@5D{#1}%
 \else\mathaccent"0365{#1}\fi}
\font\teneufm=eufm10 scaled\magstep1
\font\seveneufm=eufm7 scaled\magstep1
\font\fiveeufm=eufm5 scaled\magstep1
\newfam\eufmfam
\textfont\eufmfam=\teneufm
\scriptfont\eufmfam=\seveneufm
\scriptscriptfont\eufmfam=\fiveeufm

\csname amssym.def\endcsname
\newif{\ifcomentarios}
\comentariosfalse
\renewcommand{\theequation}{\thesection.\arabic{equation}}

\newcommand{\zerarcounters}
{
\setcounter{equation}{0}
\setcounter{theorem}{0}
}

\newcommand{\be}{\begin{equation}}
\newcommand{\ee}{\end{equation}}
\newcommand{\bma}{\begin{displaymath}}
\newcommand{\ema}{\end{displaymath}}
\newcommand{\bc}{\begin{center}}
\newcommand{\ec}{\end{center}}
\newcommand{\text}{\rm}

\newcommand{\uflex}
{{\scriptstyle {\raise 9pt\hbox{$\backslash$}\,\!\!\!\!\!\Bigg\vert}}}
\newcommand{\ncm}{\newcommand}

\ncm{\rncm}{\renewcommand}
\ncm{\id}{{\bf 1}}
\ncm{\beq}{\begin{equation}}
\ncm{\eeq}{\end{equation}}
\ncm{\ba}{\begin{array}}
\ncm{\bea}{\begin{eqnarray}}
\ncm{\beanon}{\begin{eqnarray*}}
\ncm{\ea}{\end{array}}
\ncm{\eea}{\end{eqnarray}}
\ncm{\eeanon}{\end{eqnarray*}}
\ncm{\fns}{\footnotesize}
\ncm{\setc}[1]{\setcounter{equation}{#1}}
\newcounter{eqnr}
\newenvironment{eqnarrayabc}{\stepcounter{equation}
  \setcounter{eqnr}{\value{equation}}\setc{0}
  \rncm{\theequation}{\thesection.\arabic{eqnr}\alph{equation}}
  \begin{eqnarray}}{\end{eqnarray}\setc{\value{eqnr}}}
\ncm{\eqboxabc}[3]{\newline\parbox[t]{1.5cm}{#1}\hfill
  \parbox[b]{12cm}{\begin{eqnarray*} #3\end{eqnarray*}}\hfill
   \parbox[b]{1.5cm}{\vspace{-0.0cm}\begin{eqnarrayabc}#2\end{eqnarrayabc}}\newline}
\ncm{\eqbox}[2]{\newline\parbox{1.5cm}{#1}\hfill
  \parbox{12cm}{\beanon #2\eeanon}\hfill
  \parbox{1cm}{\bea\eea}\newline}
\ncm{\nr}[1]{\parbox{1cm}{\begin{eqnarrayabc}#1\end{eqnarrayabc}}\\}

\ncm{\kal}[1]{\mbox{$\cal #1 $}}
\ncm{\mrk}[1]{\!\!\! #1 \!\!\!} 
\ncm{\qed}{\hspace*{0.4cm}\rule{0.24cm}{0.24cm}}  
\ncm{\mbold}[1]{\mbox{\boldmath $ #1 $}}   
\ncm{\bm}{\mbold}
\ncm{\str}{\stackrel}
\ncm{\sub}{\subset}
\ncm{\e}{\varepsilon}
\ncm{\ka}{\kappa}
\ncm{\inputc}[1]{\begin{center}\input{#1}\end{center}}
\ncm{\lto}{\longrightarrow}
\ncm{\x}{\times}
\ncm{\bmm}{\bm{\cal M}}
\ncm{\cp}{{\bf P}}    
\ncm{\bfp}{{\bf P}}
\ncm{\bmi}{\bm{i}}
\ncm{\bmom}{\bm{\om}}
\ncm{\bmOm}{\bm{\Om}}
\ncm{\res}{\restriction}
\ncm{\bmL}{\bm{\cal L}}
\ncm{\bmell}{\bm{\ell}}
\ncm{\bmE}{\bm{\cal E}}
\ncm{\bme}{\bm{e}}
\ncm{\bmpi}{\bm{\pi}}
\ncm{\bmr}{\bm{r}}
\ncm{\bmsigma}{\bm{\sigma}}
\ncm{\wt}{\widetilde}
\newcommand{\beaa}{\begin{eqnarray}}
\newcommand{\eeaa}{\end{eqnarray}}

\begin{document}
\input{epsf.tex}

\author{{\bf Oscar Bolina}\thanks{Supported by FAPESP under grant
97/14430-2. {\bf E-mail:} 
bolina@math.ucdavis.edu} \\
Department of Mathematics\\
University of California, Davis\\
Davis, CA 95616-8633 USA\\
}
\title{\vspace{-1in}
{\bf Intrinsic Kinematics}}
\date{}
\maketitle
\begin{abstract}
\noindent
We show how some geometric elements of the path of a particle
moving in a plane -- the osculating circle and its radius of 
curvature -- can be used to construct the parabolic
trajectory of projectiles in motion under gravity.

\noindent
{\bf Key words:} Osculating Circle, Radius of Curvature,
Parabolic Motion \hfill \break
\end{abstract}



\section{Intrinsic Equations}
\zerarcounters

The acceleration vector of a material point moving in a plane can 
be resolved in two special directions which are independent of the 
choice of the particular system of reference used to describe the 
motion. These {\it intrinsic} directions are the tangent to the
trajectory of the material point and the perpendicular to it in 
the plane of the motion \cite{Pal}. 
\newline
Fig.1a shows the situation for a particle describing an arbitrary 
trajectory in the plane. At the position {\it P} of the particle 
we have indicated the direction of the velocity vector $\vec{v}$ 
and the total acceleration vector $\vec{a}$.
\newline
The component of the acceleration tangent to the path, $a_{t}$, 
measures the rate of change of the magnitude of the velocity 
vector. The component of the acceleration normal to the path,
$a_{n}$, measures the rate of change of the direction of the 
velocity vector. 
\newline
In Fig. 1a, we have also drawn a circle of radius $\rho$ and 
center {\it O} which is tangent to the path at {\it P}. When 
this circle fits the curve just right at {\it P} it is called 
the {\it osculating circle} of the path at that point. The 
osculating circle is very helpful in determining the component 
of the acceleration normal to the path of the particle. If we 
imagine that when the particle is at {\it P}, instead of following 
its real path, it describes a uniform motion around the osculating 
circle itself, the component of the acceleration normal to the 
path becomes the centripetal acceleration in this motion, having 
magnitude $a_{n}=v^{2}/{\rho}$ in direction of the radius {\it PO} 
of the osculating circle \cite{{Pal},{BW}}. 
\newline
Now note that in Fig.1a we have represented the total acceleration 
vector in the direction of the chord {\it PQ} of the osculating 
circle, making an angle $\phi$ with the radius {\it PO}. Since
$a_{n}=a \cos \phi$ we also have 
\beq \label{1}
\frac{v^{2}}{\rho}=a\cos\phi. 
\eeq
The magnitude of the total acceleration of the particle in (\ref{1}) 
can be related to yet another geometric element of the osculating 
circle, namely, the length of the chord {\it PQ} between the particle
and the osculating circle, in the direction of the total acceleration
vector. From Fig. 1a we see that this length is $C=2\rho \cos\phi$.
Substituting this value for $\phi$ into (\ref{1}) yields:
\beq 
C=\frac{2 v^{2}}{a} \label{2}.
\eeq


\section{Projectile Motion}
\zerarcounters

Relation (\ref{2}) finds an interesting application in the study 
of projectile motion under gravity \cite{Y}. In Fig. 1b we have
represented the parabolic trajectory described by a projectile
fired with velocity $\vec{v_{0}}$ at an angle $\theta$ to the
horizontal. Suppose that when the projectile is at {\it P} its 
velocity vector $\vec{v}$ makes an angle $\beta$ to the horizontal. 
Seeing that the horizontal projection of the motion is uniform, 
the equality $v\cos\beta=v_{0}\cos\theta$ holds at {\it P}. The 
acceleration in the direction of the chord {\it PQ} is {\it g} 
due to gravity. Thus Eq. (\ref{2}) becomes
\beq\label{Ch}
C_{\beta}=\frac{2 v_{0}^{2}\cos^{2}\theta}{g \cos^{2}\beta}.
\eeq
Formula (\ref{Ch}) allows us to construct the parabolic motion of 
the projectile from the intrinsic elements developed above \cite{D}. 
First we note that when $\beta=0$ the particle reaches the 
{\it vertex V} of the parabola. The length of the chord {\it PQ} 
in this position is 
\beq \label{P}
C_{(\beta=0)}=2 p =\frac{2 v_{0}^{2}\cos^{2}\theta}{g}.
\eeq
The above relationship determines a length {\it p} which is 
the distance between the {\it focus} and the {\it directrix 
line} of the parabola. This distance is the basis for the 
construction of the parabola, as we will see in the next section, 
since the defining property of a parabola is that any point on it 
is equidistant from the focus and the directrix.


\section{The Parabola}
\zerarcounters

We begin the construction of the parabola by tracing the line {\it PH} 
normal to the path along the radius of the osculating circle (But note 
that {\it H} is {\it not} the center of the circle), and {\it PG} along 
the horizontal, as shown in Fig.1b. The {\it axis} of the parabola is 
the vertical line through {\it H} and {\it G} when $GH=p$. 
\newline
To locate the {\it focus} we invoke the {\it reflective property} 
of the parabola, according to which any {\it light ray} ({\it PQ})
parallel to the axis and incident on the parabola is reflected to 
the focus. The ray {\it PQ} incides on the parabola at {\it P} making
an angle $\beta$ with respect to {\it PH}, and is refleted to {\it F}
in such a way that the angle of reflection equals
the angle of incidence, or $\widehat{HPF}=\beta$.
\newline
It follows from simple trigonometric relations in the triangles 
{\it PGF} and {\it PGH} that
\[
PH =2 PF \cos \beta \;\;\;\;\;\; {\rm and} \;\;\;\;\;\;
p= PH \cos \beta.
\]
Eliminating {\it PH} among the above equations we get
the following expression for the distance between of 
the particle to the focus of the parabola
\beq \label{Fo}
PF= \frac{p}{2 \cos^{2} \beta}.
\eeq
From (\ref{Fo}) we see that the vertex of the parabola is 
a point on its axis at a distance $p/2$ from the focus.
The distance (\ref{Fo}) is related the length (\ref{Ch}) 
of the chord {\it PQ} at any point of the path of the
particle by $C_{\beta}=4 PF$. 
\newline
These elements suffice to construct the parabola (See \cite{Coe}
for the more usual analytical description). 
\newline
Finally, we mention that two important parameters pertaining 
to a more physical analysis of the projectile motion are the 
total horizontal distance (or range) and the maximum height 
attained by the projectile in the case it returns to the same 
horizontal it was launched at. We work out the expression for 
the range here, and leave it to the reader to figure out how 
to use our analysis to determine the maximum height.
\newline
The range {\it R} is just twice the horizontal distance 
{\it PG} when {\it P} is the launching point of the 
projectile, and {\it G} is a point on same the horizontal
from {\it P}. In this case {\it PG} forms an angle 
${\pi}/2-\theta$ with the corresponding line {\it PH}, 
and we obtain 
\beq
R=2 p \tan \theta=\frac{v_{0}^{2} \sin 2 \theta}{g}.
\eeq


\newpage
\begin{figure}
\centerline{
\epsfbox{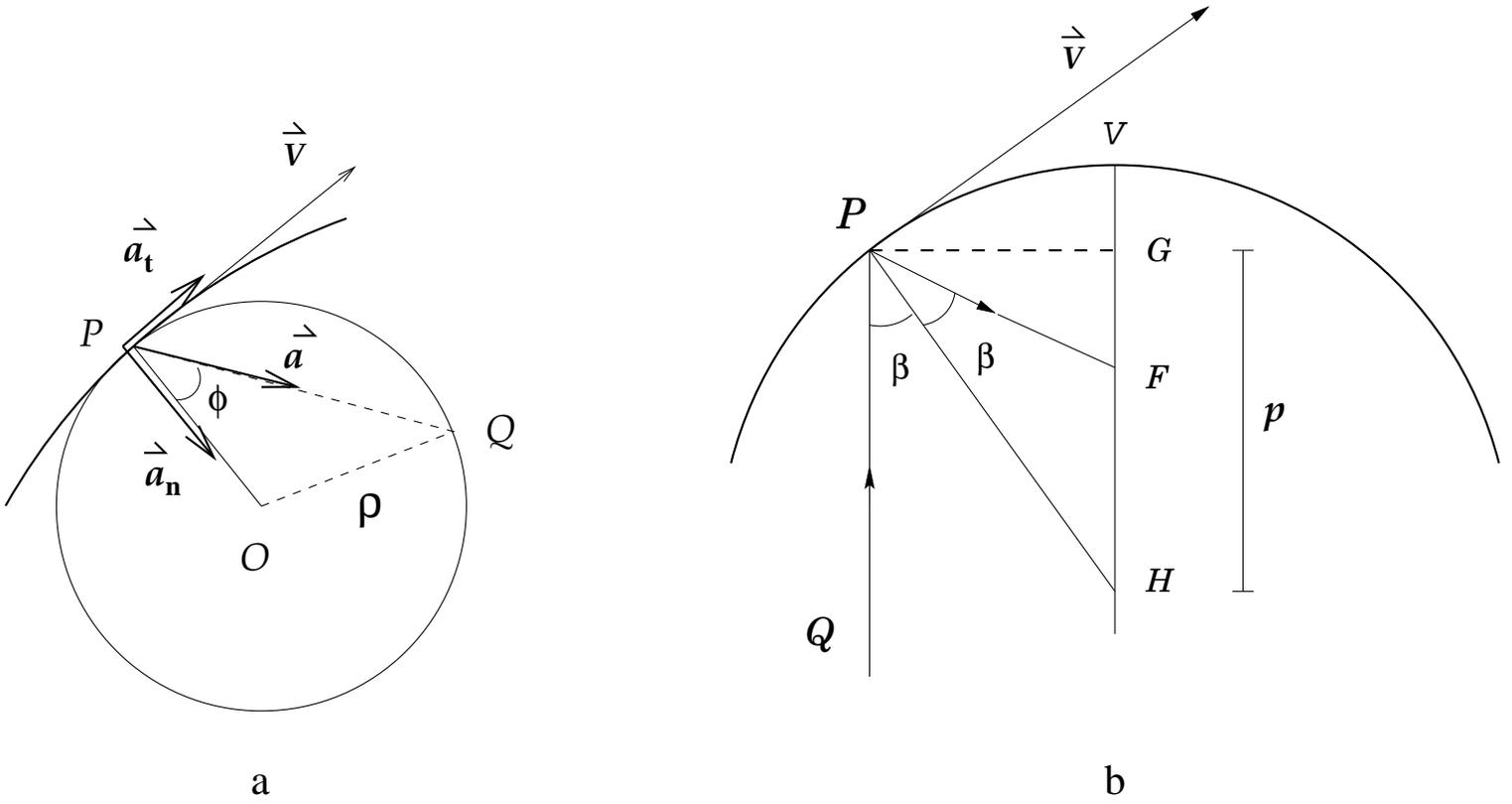}}
\vskip 5 cm
\caption{(a) Arbitrary plane motion, and (b)
parabolic motion.}
\end{figure}

\end{document}